\documentclass[9pt,conference]{IEEEtran}
\IEEEoverridecommandlockouts
\usepackage{cite}
\usepackage{amsmath,amssymb,amsfonts,mathtools}
\usepackage[caption=false,font=footnotesize,labelfont=sf,textfont=sf]{subfig}
\usepackage{subcaption}
\usepackage{float}
\usepackage{algorithmic}
\usepackage{graphicx}
\usepackage{textcomp}
\usepackage{xcolor}
\usepackage{multicol,multirow}
\usepackage{array} 
\usepackage{hyperref} 

\renewcommand{\IEEEbibitemsep}{0.5em} 

\def\BibTeX{{\rm B\kern-.05em{\sc i\kern-.025em b}\kern-.08em
    T\kern-.1667em\lower.7ex\hbox{E}\kern-.125emX}}
\begin{document}

\title{MIMII-Gen: Generative Modeling Approach for \\Simulated Evaluation of Anomalous Sound Detection System 
}

\author{\IEEEauthorblockN{
Harsh Purohit, Tomoya Nishida, Kota Dohi, Takashi Endo, and Yohei Kawaguchi}
\IEEEauthorblockA{\textit{R\&D Group, Hitachi Ltd.}
\\
Tokyo, Japan\\
\small
\small\texttt{[harsh\_pramodbhai.purohit.yf,
yohei.kawaguchi.xk]@hitachi.com} }}
\maketitle

\maketitle

\begin{abstract}
Insufficient recordings and the scarcity of anomalies present significant challenges in developing and validating robust anomaly detection systems for machine sounds. To address these limitations, we propose a novel approach for generating diverse anomalies in machine sound using a latent diffusion-based model that integrates an encoder-decoder framework. Our method utilizes the Flan-T5 model to encode captions derived from audio file metadata, enabling conditional generation through a carefully designed U-Net architecture. This approach aids our model in generating audio signals within the EnCodec latent space, ensuring high contextual relevance and quality. We objectively evaluated the quality of our generated sounds using the Fréchet Audio Distance (FAD) score and other metrics, demonstrating that our approach surpasses existing models in generating reliable machine audio that closely resembles actual abnormal conditions. The evaluation of the anomaly detection system using our generated data revealed a strong correlation, with the area under the curve (AUC) score differing by 4.8\% from the original, validating the effectiveness of our generated data. These results demonstrate the potential of our approach to enhance the evaluation and robustness of anomaly detection systems across varied and previously unseen conditions. 
Audio samples can be found at \url{https://hpworkhub.github.io/MIMII-Gen.github.io/}.


\end{abstract}

\vspace*{0.05cm}
\section{Introduction}

Anomaly detection in machine operations is critical, as even minor deviations in machine sounds can indicate significant faults in industrial applications \cite{kawaguchi2021description}. These systems must effectively identify irregularities across various machines, operational changes, and recording environments. However, evaluating anomaly detection models poses a significant challenge due to the scarcity of recorded data, particularly abnormal samples, which are often difficult to obtain in real-world settings. This limitation hinders the development of robust and accurate detection systems.

To address this, we propose an approach that leverages audio generation to create abnormal samples, enabling a simulated evaluation of anomaly detection systems. Unlike current audio generation methods such as Tango, AudioLDM, MusicLM \cite{agostinelli2023musiclm, chen2022resgrad, huang2023make, liu2023audioldm, ghosal2023text, yang2023diffsound} have shown remarkable success in domains like speech and music, generating high-quality machine audio remains an under-explored area. Existing techniques fail to accurately replicate the fine-grained audio differences and complexities of machine sounds due to different operational environments. Addressing this gap, our research named as MIMII-Gen seeks to advance machine sound generation similar to recorded MIMII-DG data \cite{dohi2022mimii}, enabling practical applications, particularly in evaluating anomaly detection systems.

We develop and train a latent diffusion model that integrates an encoder-decoder approach of EnCodec \cite{defossez2022high} for converting between audio clips and their latent representations. To list our contributions, (i) We enhance the descriptive quality of weak metadata associated with sound clips by converting them into rich, human-like captions. (ii) We have carefully designed the U-Net for improved quality while guiding the generation through conditional embeddings of captions from Flan-T5 \cite{chung2024scaling}. The results, outlined in section \ref{section:results}, show that our approach outperforms current baseline generation models by producing reliable machine audio samples with subtle, fine-grained variations essential for industrial applications such as anomaly detection. (iii) We validate the effectiveness of our generated data by assessing the performance of anomaly detection system using standard evaluation metrics.

The remainder of this paper is organized as follows: Section \ref{section:proposedapproach} details our proposed method, including modeling and training; Section \ref{section:Experiments} describes the dataset and experimental setup; and Section \ref{section:results} presents (i) evaluation and comparison with baseline models using FAD and other metrics as well as (ii) evaluation of anomaly detection system using generated data.

\begin{figure*}[ht]
	\begin{center}
        \includegraphics[width=0.8\hsize,height=0.4\hsize,clip]{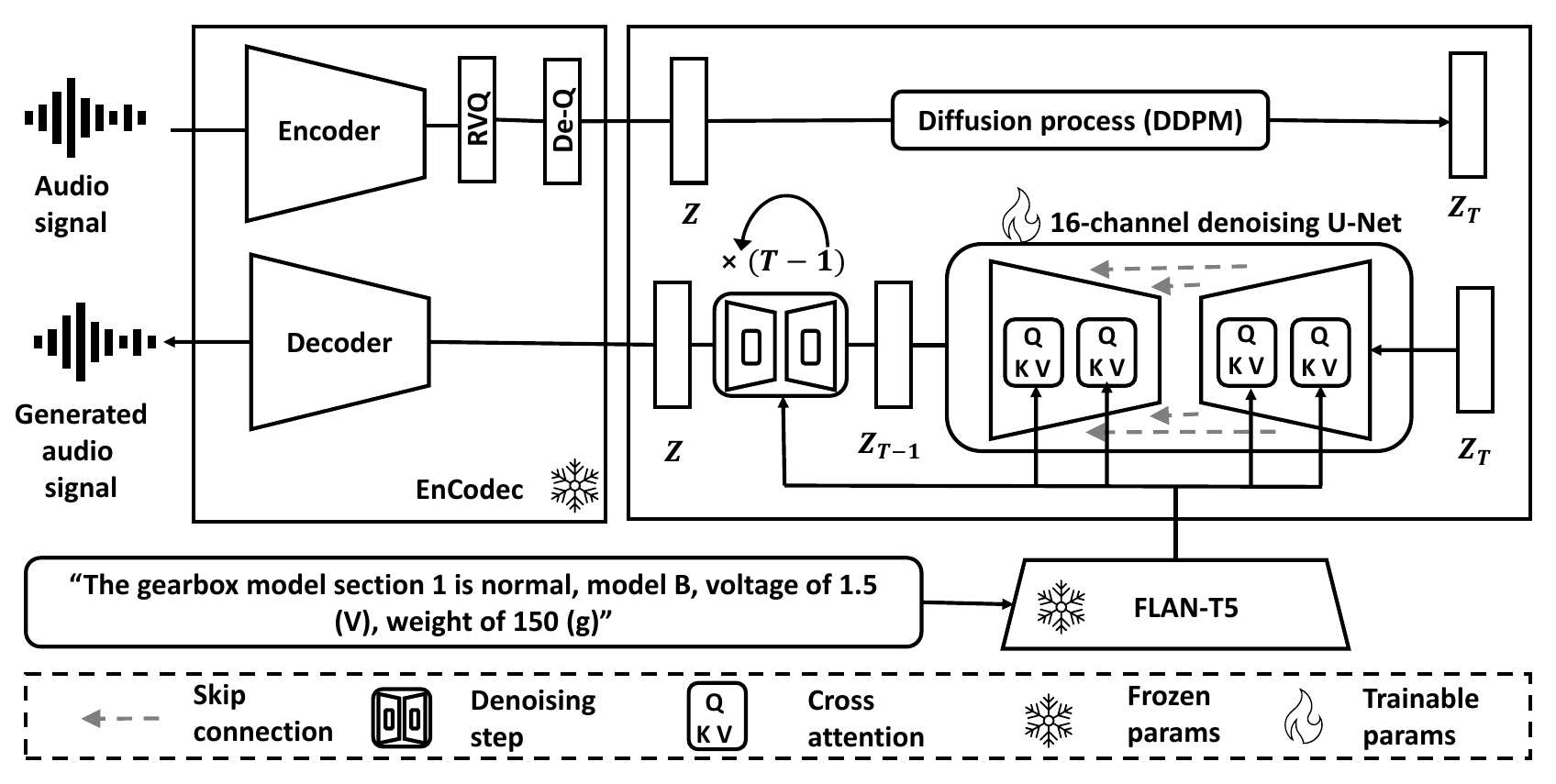}
	\caption{Block diagram of proposed approach for machine sound generation.}
	\label{fig:block_d}
	\end{center}
\end{figure*}

\section{Related work}

In this section, we review the recent advancements in audio generation using diffusion-based models and unsupervised anomaly detection in machine sounds.

\subsection{Text-to-Audio Generation Using Diffusion Models}
Text-to-audio (TTA) generation has garnered considerable attention, with approaches like AudioGen \cite{kreuk2022audiogen} focusing on learning audio representations by leveraging paired audio-text data to overcome the challenges of data scarcity and quality variability. AudioGen employs a Transformer-based decoder to generate discrete tokens that represent compressed waveform features. By implementing data augmentation techniques, such as mixing audio samples and distilling language descriptions into simplified labels, AudioGen increases the diversity of training data. However, this comes at the cost of losing intricate spatial and temporal relationships in the text descriptions, which can impact the fidelity and contextual richness of the generated audio.
On the other hand, diffusion models have become a dominant framework for generative tasks, including text-to-audio conversion.
Diffsound\cite{yang2023diffsound}, uses a non-autoregressive decoder based on discrete diffusion model to generate audio using text by refining mel-spectrogram tokens through iterative steps rather than sequential predictions typical of autoregressive decoders.

Generation approaches like AudioLDM \cite{liu2024audioldm}, Make-An-Audio \cite{huang2023make2}, and Tango \cite{ghosal2023text} typically employ pre-trained text encoders (e.g., CLAP \cite{elizalde2023clap}, T5) and VAEs to extract text embeddings and latent audio features. Using a latent diffusion model (LDM) architecture, these systems generate audio latent features conditioned on text inputs, which are subsequently transformed into mel-spectrograms and waveforms using VAEs and neural vocoders. 

Despite these advances, current TTA models often fall short when applied to machine sound generation, where the complexity of acoustic environments and subtle variations in sound are critical. The existing methods primarily focus on speech and music, with limited exploration of industrial machine sounds. This gap underscores the need for specialized generative models tailored to the unique challenges of machine audio.

\subsection{Unsupervised Anomaly Detection in Machine Sounds}
Anomalous sound detection (ASD) \cite{koizumi2017optimizing,kawaguchi2019anomaly,purohit2020deep,suefusa2020anomalous,purohit2022hierarchical} aims to identify deviations from normal operational sounds in machines, a task complicated by the rarity and variability of anomalous events. Traditional ASD approaches often rely on labeled anomalous data, which is scarce in real-world applications, limiting their ability to generalize to new or unseen conditions. Consequently, unsupervised ASD, which trains only on normal sounds, has emerged as a viable yet challenging alternative.

The DCASE-2023 Challenge Task-2 introduced a first-shot (FS) approach to unsupervised ASD, targeting the detection of anomalies in machine types not seen during training \cite{dohi2023description}. 
However, unsupervised ASD methods struggle to adapt to first-shot scenarios due to a lack of diverse training data encompassing unseen machine types and operational conditions. Limited anomalous data for evaluation also hinders adaptability and reliability in real-world scenarios where anomalies vary widely.



Recent work by Zhang et al. \cite{zhang2024first} uses audio generation of anomalous sounds for training and enhancing anomaly detection systems, whereas our approach focuses on generating anomalies to evaluate the robustness and effectiveness of existing anomaly detection system. Our approach combines generative modeling with EnCodec and Flan-T5 embeddings to produce machine audio that captures subtle variations crucial for anomaly detection. Tailored for industrial acoustic environments, it generates diverse audio data for anomaly detection evaluation when real-world data is scarce.


\section{Proposed approach}
\label{section:proposedapproach}

 
    Our approach employs a condition-based latent diffusion model to generate machine sounds under various operational and environmental conditions. The crucial parts of the generation model as well as diffusion process are explained below.

\subsection{Overview of Diffusion Models}

Diffusion models are probabilistic generative models \cite{rombach2022high} that learn the data distribution \( p(x) \) by progressively denoising a normally distributed variable. The diffusion process involves learning the reverse of a fixed Markov chain of length \( T \), where the forward process adds noise step-by-step, and the reverse process denoises iteratively. This approach can be conceptualized as a sequence of denoising autoencoders, denoted as \(\epsilon_\theta\left(x_t, t\right)\) is a noisy version of the input \( x \) at time step \( t \).

The objective function for diffusion models is expressed as:

\begin{equation}
L_{DM} = \mathbb{E}_{x, \epsilon \sim \mathcal{N}(0, 1), t} \left[\left\|\epsilon - \epsilon_\theta\left(x_t, t\right)\right\|_2^2\right],
\end{equation}

where \( t \) is uniformly sampled from \(\{1, \ldots, T\}\). This formulation mirrors denoising score matching, enabling the model to predict clean data from noisy observations effectively.
Instead of high-dimensional data space, diffusion models can operate in latent space using low-dimensional representations from an encoder, thus reducing computations and focusing on semantically relevant aspects of the data.

%


The training objective within the latent diffusion framework is defined as:

\begin{equation}
L_{LDM} = \mathbb{E}_{\mathcal{E}(x), \epsilon \sim \mathcal{N}(0, 1), t} \left[\left\|\epsilon - \epsilon_\theta\left(z_t, t\right)\right\|_2^2\right],
\end{equation}

where \( z_t \) represents the noisy latent variable derived from the encoder, and the reverse process is modeled through a time-conditional U-Net backbone as shown in Fig. \ref{fig:block_d}.


Diffusion models can also model conditional distributions  of the form \( p(z|y) \) by projecting the conditioning variable $y$ to representations $\tau_\theta(y)$ through encoder $\tau_\theta$. The loss function for the conditional latent diffusion model is expressed as follows:

\begin{equation}
L_{LDM} = \mathbb{E}_{\mathcal{E}(x), y, \epsilon \sim \mathcal{N}(0, 1), t} \left[\left\|\epsilon - \epsilon_\theta\left(z_t, t, \tau_\theta(y)\right)\right\|_2^2\right],
\end{equation}

where both the latent encoder \( \epsilon_\theta \) and the conditioning encoder \( \tau_\theta\) can be jointly optimized according to the loss function. 



 \subsection{Model architecture}

 As shown in Fig.\ref{fig:block_d}, we generate captions from the metadata of audio files, which describe operational settings, environmental conditions, anomaly types, and machine models. These captions are then encoded into dense vector representations that serve as condition vectors for the diffusion model to generate realistic machine sounds across combinations of different conditions. EnCodec is used to obtain latent space representations of audio signals, capturing essential features in a compressed form. The diffusion model is trained using a Denoising Diffusion Probabilistic Model (DDPM) framework utilizing a wide-channel denoising U-Net where the condition embeddings are combined with audio representations.

\subsubsection{Text condition encoding}
We utilize the Flan-T5 model\cite{chung2024scaling} as a text encoder to obtain condition embeddings from generated captions. These encoded caption embeddings from 768-dimensional projection layer of the text encoder are then used as conditions for the U-Net model.
\subsubsection{Generation of latent space}
Many audio generation approaches, such as AudioLDM and Tango, rely on a combination of VAE and vocoder architectures. These methods apply diffusion to spectrogram representations and require additional training of a VAE \cite{liu2023audioldm} and a separate vocoder (e.g., HiFi-GAN \cite{kong2020hifi}) to reconstruct waveforms from generated spectrograms. In contrast, we use EnCodec, an off-the-shelf VQ-GAN model, known for its effective performance in audio generation similar to AudioJourney\cite{michaels2024audio}. The EnCodec model comprises an encoder, a vector quantizer, and a decoder, enabling efficient audio generation thus reducing the overall model size and complexity.

In Encodec, given a continuous latent represention that comes out of the encoder, residual vector quantization (RVQ) is applied to convert it into discrete set of indexes of codebooks. The dimension $N$ of latent representation thus becomes $N_q$ which is equal to the number of codebooks
selected. Variable bandwidth training is performed in Encodec where the authors select randomly a number of codebooks as a multiple of 4, i.e. corresponding to a bandwidth 1.5, 3, 6, 12 or 24 kbps at 24 kHz. We experimented with these different bandwidth values and selected 24 kbps for our use case based on generation quality. This discrete representation can changed again to a vector by summing the corresponding codebook
entries, which is done just before going into the decoder by the dequantization block shown as De-Q in Fig.\ref{fig:block_d}. We use this continuous latent vector as input for diffusion process during training. During inference, the sampled latent vector is then directly given as input to the decoder for generating the audio clip.
  
  \subsubsection{U-Net design and noise scheduling}
  
In order to effectively utilize the latent space from the encoder of Encodec model, we use a wide channel U-Net in the diffusion process, that is designed for a 16-channel input rather than the typical one or three channels seen in many audio generation methods. To counter the minimal variance within the 16-dimensional latent encodings from encoder, we reshape the latent vectors from a single-channel $128\times 750$ format to a $16$-channel $8 \times 750$ format. With this new structure, the convolutional blocks encompass the entire latent representation within their receptive field without losing detail, resulting in higher fidelity audio generation. This reshaping allowed each channel to be separately normalized to zero mean and unit variance, aiding the U-Net in learning the noise distribution, $N(0, I)$. The approach of reshaping the encoder output is inspired from Audiojourney although the resulting dimension due to reshaping, number of channels and U-Net design are different due to our specific data characteristics. Post-generation, these transformations are fully reversible, enabling decoding back into waveforms using decoder of Encodec model. Utilizing cross-attention rather than embedding addition in the U-Net architecture allows original audio embeddings to remain unchanged throughout each layer, thereby retaining their integrity and enhancing conditional guidance.

The generated samples are used to evaluate generation quality and as input into an auto-encoder based anomaly detection system \cite{dohi2023description}. 

\begin{table*}[h]
\centering
\caption{Generated and ground truth audio clips. Listen to them at: \url{https://hpworkhub.github.io/MIMII-Gen.github.io/}}
\renewcommand{\arraystretch}{1.5} 
\setlength{\tabcolsep}{6pt}       
\begin{tabular}{m{2cm} m{3.5cm} m{5cm} m{5cm} }
\hline
\textbf{Metadata} & \textbf{\hspace{1cm} Caption} & \textbf{\hspace{1.8cm} Ground Truth } & \textbf{\hspace{1.8cm}  Generated }\\
\hline
\raggedright{Bearing, anomaly, axis damage, velocity of 24 krpm, location "A"} & A bearing operating on velocity of 24 krpm with anomaly due to axis damage at location A & 
\vspace{0.2pt}\raisebox{-0.5\height}{\includegraphics[width=5cm]{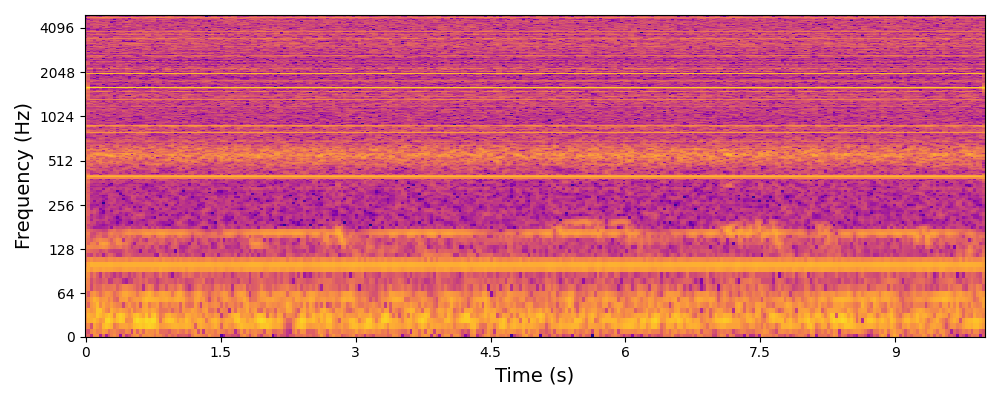}}\vspace{0.2pt} & 
\vspace{0.2pt}\raisebox{-0.5\height}{\includegraphics[width=5cm]{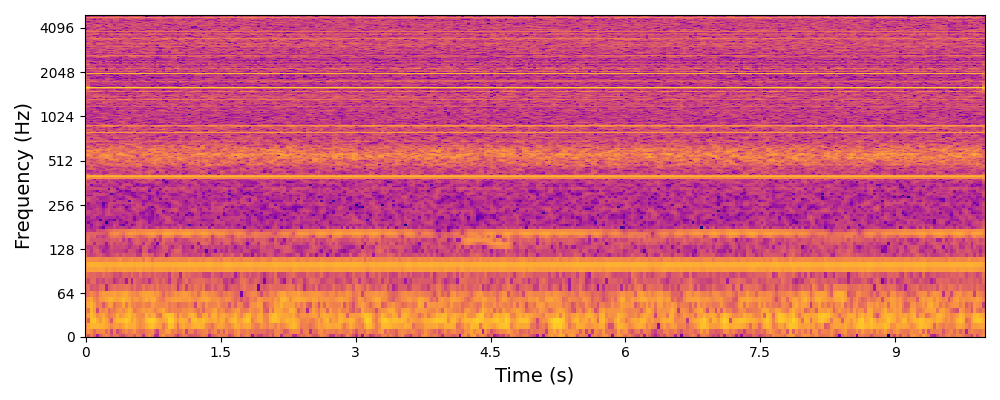}}\vspace{0.2pt} \\
\hline
\raggedright{Gearbox, anomaly, damage type 2, model B, voltage of 2.3 (V), weight of 0 (g)} & \raggedright{A gearbox model B operating on voltage of 2.3 (V) and weight of 0 (g) with anomaly due to damage type 2} & 
\vspace{0.2pt}\raisebox{-0.5\height}{\includegraphics[width=5cm]{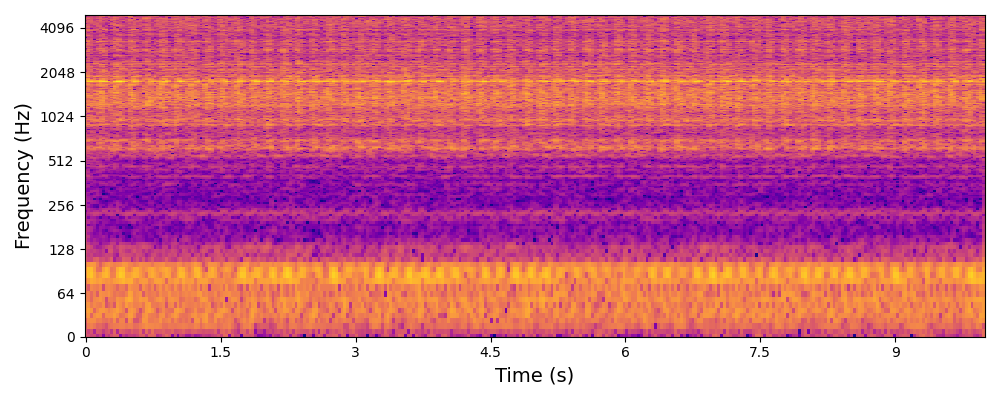}}\vspace{0.2pt} & 
\vspace{0.2pt}\raisebox{-0.5\height}{\includegraphics[width=5cm]{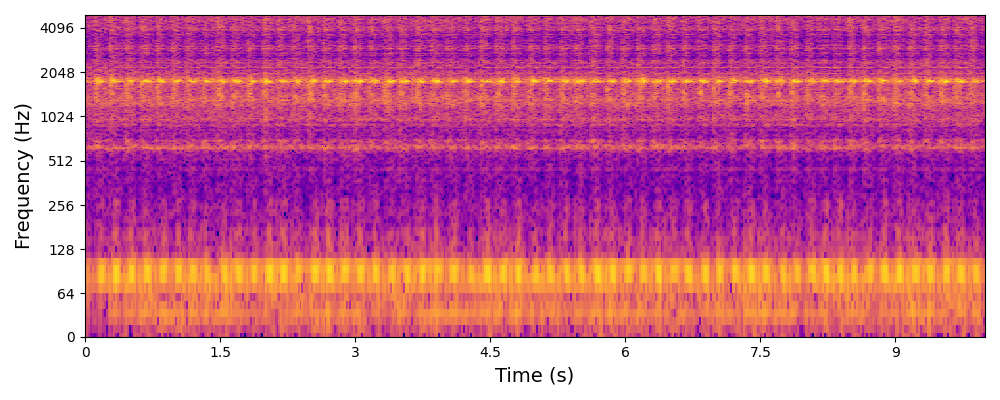}}\vspace{0.2pt} \\
\hline
\raggedright{Fan, model, anomaly, over voltage} & \raggedright{A fan model is running on over voltage with anomaly} & 
\vspace{0.2pt}\raisebox{-0.5\height}{\includegraphics[width=5cm]{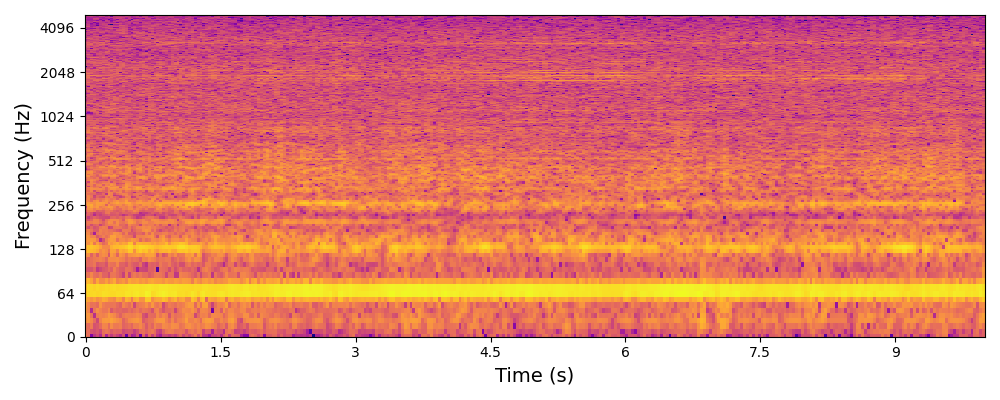}}\vspace{0.2pt} & 
\vspace{0.2pt}\raisebox{-0.5\height}{\includegraphics[width=5cm]{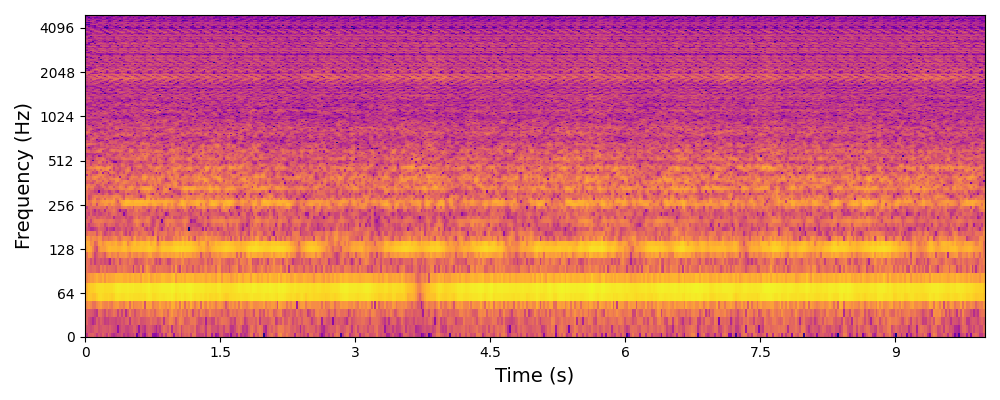}}\vspace{0.2pt} \\
\hline
\raggedright{Slider, ball-type, anomaly, damage, velocity 1000.0 (mm/s), acceleration 0.3} & \raggedright{A ball-type slider operating on velocity of 1000.0 (mm/s) and an acceleration of 0.3 is with an anomaly due to damage} & 
\vspace{0.2pt}\raisebox{-0.5\height}{\includegraphics[width=5cm]{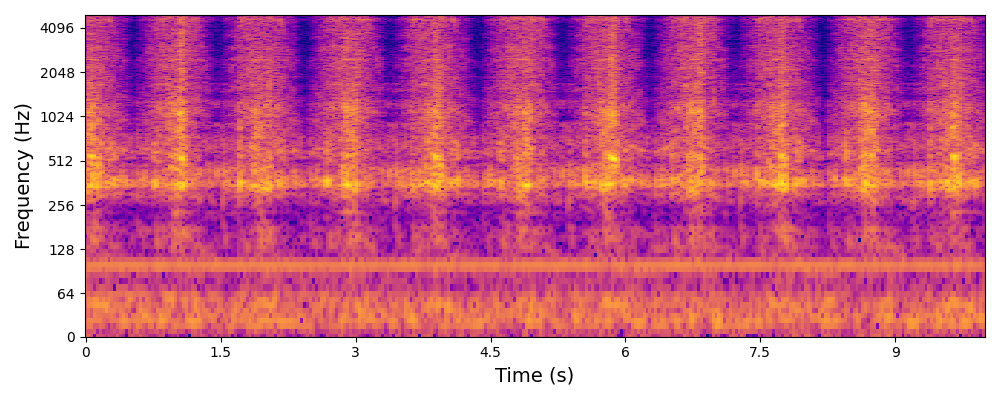}}\vspace{0.2pt} & 
\vspace{0.2pt}\raisebox{-0.5\height}{\includegraphics[width=5cm]{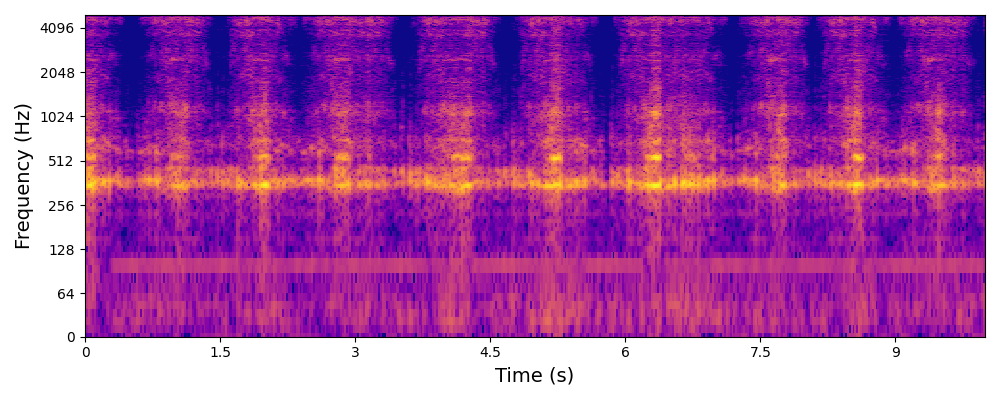}}\vspace{0.2pt} \\
\hline
\raggedright{Valve, anomaly, contamination, moving pattern 1, surroundings is open} & \raggedright{A valve of moving pattern 1 in open surroundings with anomaly due to contamination} & 
\vspace{0.2pt}\raisebox{-0.5\height}{\includegraphics[width=5cm]{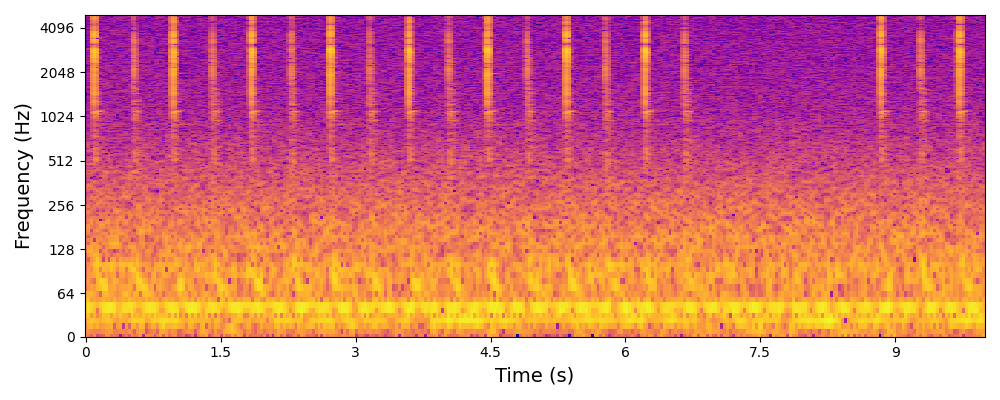}}\vspace{0.2pt} & 
\vspace{0.2pt}\raisebox{-0.5\height}{\includegraphics[width=5cm]{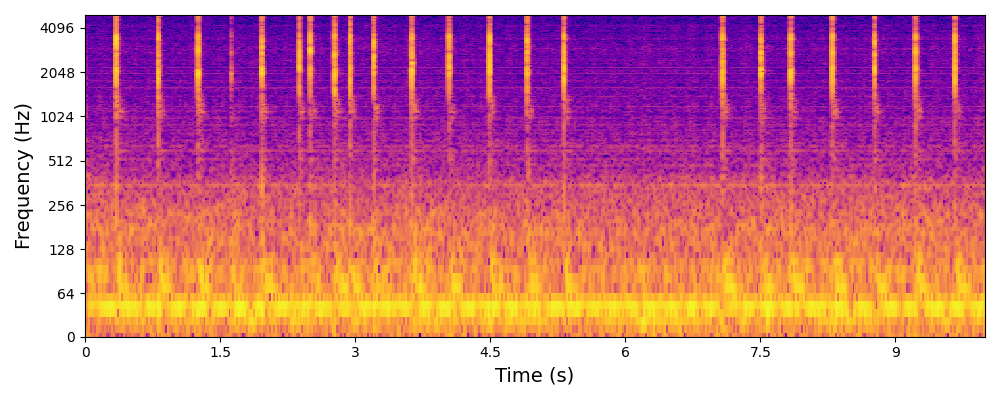}}\vspace{0.2pt} \\
\hline
\end{tabular}
\label{tab:spectrograms}
\end{table*}

\section{Experiments}
\label{section:Experiments}
 
\subsection{Dataset}
The data used for this research is originally from MIMII-DG dataset \cite{dohi2022mimii} that has sounds of various machines recorded in different operating conditions. This dataset is then used for anomaly detection Task-2 in DCASE \cite{dohi2023description} which is composed of three parts: the development dataset, an additional training dataset, and an evaluation dataset. The development dataset includes seven machine types (fan, gearbox, bearing, slide rail, ToyCar, ToyTrain, valve), with each type featuring one section that comprises complete training and test data. Various attributes related to operational and environmental conditions as well as model types of the machines are available as metadata. 
We used recordings of the audio sounds of five machines (fan, gearbox, bearing, slide rail, valve) from MIMII-DG corresponding to this development dataset part for training and evaluating our generation model. Each audio recording is a single-channel file lasting about 10 seconds with a sampling rate of 16 kHz.


\begin{table}[H]
\centering
\caption{Counts of audio samples for each machine type in training and validation datasets}
\begin{tabular}{p{1.2cm} p{3cm}<{\raggedleft} p{3cm}<{\raggedleft}}
\hline
\textbf{Machine Type}        & \textbf{Training Samples} & \textbf{Validation Samples} \\ \hline
Bearing                      & Normal: 6381              & Normal: 1613               \\
                             & Anomalous: 637            & Anomalous: 151             \\ \hline
Gearbox                      & Normal: 5863              & Normal: 1485               \\
                             & Anomalous: 778            & Anomalous: 192             \\ \hline
Fan                          & Normal: 5961              & Normal: 1458               \\
                             & Anomalous: 850            & Anomalous: 202             \\ \hline
Slide rail                   & Normal: 5983              & Normal: 1536               \\
                             & Anomalous: 1036           & Anomalous: 286             \\ \hline
Valve                        & Normal: 7134              & Normal: 1707               \\
                             & Anomalous: 523            & Anomalous: 157             \\ \hline
\end{tabular}
\label{tab:datacount}
\end{table}

\subsection{Experimental setup}

\subsubsection{Audio generation}

In order to train the diffusion model, total of 35,146 training samples from all machines are used and 8,787 samples are used for evaluation of the generation quality. Table \ref{tab:datacount} shows the distribution of samples from each machine type. 

The metadata associated with all the recorded audio clips are given as input to the T5-large \cite{raffel2020exploring} model to obtain descriptive captions that are saved in order to be used later for audio generation. Table \ref{tab:spectrograms} shows examples of the metadata and captions of the dataset.
These captions are then encoded into $768-d$ embedding vectors using another model Flan-T5 as shown in Fig. \ref{fig:block_d} to give as input condition during training and inference from diffusion model. During training, the parameters of Encodec and the Flan-T5 model are frozen, only the denoising U-Net is trained.

\subsubsection{Anomaly detection}

In order to train the anomaly detection system, we used $990$ normal audio clips from each machine type for training. Two evaluation datasets are created, one set contains the originally recorded $50$ normal and $50$ anomalous clips, while other set consists of $50$ original normal and $50$ generated anomalous clips for all $5$ machine types. The anomaly detection system is same as the auto-encoder trained as baseline for task-2 of DCASE\cite{dohi2023description}

\section{Results and discussion}
\label{section:results}

We employ four objective metrics to evaluate our generated audio: Frechet Audio Distance (FAD) \cite{kilgour2018fr, gui2024adapting}, calculated using embeddings extracted by VGGish \cite{chen2020vggsound}; Kullback-Leibler divergence (KLpasst) between the outputs of PaSST \cite{koutini2021efficient}, an audio classification model; Inception Score (ISpasst) \cite{salimans2016improved}, which is also based on the outputs of PaSST; and the CLAP score. A lower FAD indicates higher audio quality for the generated samples. The KL divergence assesses the semantic similarity between generated audio and reference ground truth audio. The IS measures the diversity of the generated samples, while the CLAP score evaluates how closely the generated audio aligns with the provided textual description. We calculate the CLAP scores for two cases, (i) for original audio and caption (ii) for generated audio and the caption. The CLAP scores for both the cases should be almost same if the generated audio is similar to the original. 

Table \ref{tab:fad} presents the performance of the generation models across the evaluated metrics. The CLAP scores for the original (i) and generated (ii) cases are separated by a hyphen in the table. The results indicate that our approach outperforms the Tango baseline, which relies on AudioLDM with a pretrained VAE and vocoder. This performance gap is likely due to the vocoder’s limited generalization to non-speech audio. Additionally, our approach utilizes a 16-channel input and a wide-channel U-Net, which effectively captures the latent representations within its receptive field, enhancing denoising capabilities and resulting in improved generation quality.

\begin{table}[htbp]
\caption{FAD and other scores  for conditional audio generation.}
\centering
\begin{tabular}{ c c c c c }
\hline
\textbf{Models} & \textbf{FAD} $\downarrow$ & \textbf{KLpasst} $\downarrow$ & \textbf{ISpasst} $\uparrow$  & \textbf{CLAP score} $\uparrow$\\
\hline
Tango & 6.88 & 1.74 & 2.57 & 0.15-0.10 \\
\hline
Our approach & 5.43 & 1.22 & 3.72 & 0.15-0.14 \\
\hline
\end{tabular}

\label{tab:fad}
\end{table}
Spectrogram provides a visual representation of the audio data
highlighting the unique patterns associated with each machine type's sounds. Table \ref{tab:spectrograms} shows that spectrograms of the generated audio samples for given captions as well as the ground truth audio clips follow similar patterns. We could also successfully generate the audio samples for combinations of different conditions which were not seen during training of the diffusion model.

       


        

The anomaly detection system is then evaluated on this generated anomalous data. Table \ref{tab:anomaly} shows the AUC scores obtained on both the evaluation datasets, i.e. originally recorded as well as the generated anomalous audio clips. The AUC scores for generated data have an average difference of $4.8\%$ and are correlated to that of the original anomalous data for all machines.

\begin{table}[H]
\caption{AUC scores for all machine types}
\centering
\begin{tabular}{ c c c }
\hline
\textbf{Machine type} & \textbf{Original data} & \textbf{Generated data} \\
\hline
Bearing & 0.5468
 & 0.5916
 \\
\hline
Gearbox & 0.6920
 & 0.7607
 \\
\hline
 Fan & 0.9496
 & 0.9713
 \\
\hline
Slide rail & 0.5588
 & 0.6132
 \\
\hline
Valve & 0.5271
 & 0.5517
 \\
\hline
\end{tabular}

\label{tab:anomaly}
\end{table}

\section{Conclusion}

We presented a method for generating high-quality machine audio with fine-grained variations using metadata, demonstrating strong alignment with real recordings. Our approach effectively generates anomalous data across various conditions, enhancing downstream task performance. Leveraging an EnCodec-based approach over VAE and vocoder methods, we achieve superior performance with a simplified pipeline. The anomaly detection system tested on our generated audio shows only a $4.8$\% AUC deviation from its performance on original data, underscoring our method's practical value in industrial scenarios.



\vfill\pagebreak
\bibliographystyle{IEEEtran}
{\renewcommand{\IEEEbibitemsep}{0.6em} 
 \renewcommand{\baselinestretch}{1.18} 
 \fontsize{10pt}{12pt}\selectfont 

\bibliography{bibliography.bib}

\end{document}